\documentstyle{aipproc}
\input psfig
\newcommand{\pom}{I\!\! P}
\begin{document}
\pagestyle{plain}
\pagenumbering{arabic}
%
%
\vglue -1cm
\title{Diffraction:
Results and Conclusions\footnote{\normalsize
Presented at ``Diffractive Physics, 
LAFEX International School on High Energy Physics
(LISHEP-98), Rio de Janeiro, Brazil, 10-20 February 1998."}}
\author{Konstantin Goulianos}
\address{The Rockefeller University,
New York, NY 10021, U.S.A.}
\maketitle
\begin{abstract}
We summarize the main features of available 
experimental results on soft and hard diffraction 
and draw conclusions about the nature of the pomeron.
\end{abstract}
\section*{INTRODUCTION}
The recent results on hard diffraction from 
HERA and the Tevatron have  caused a flurry of theoretical activity.
A variety of phenomenological models exist, which 
have been successful in describing certain aspects of the data. 
However, a QCD-based 
theoretical description of diffraction  
is still lacking. This is not surprising, since diffraction
invariably involves non-perturbative effects associated with the 
formation of rapidity gaps.
The interplay between soft and hard processes in hard diffraction 
is of particular theoretical importance due to its 
potential for elucidating 
the transition from perturbative to nonperturbative QCD.
In this paper,
we summarize the main features of the available results on soft and 
hard diffraction and draw conclusions about the nature of the pomeron, 
which is presumed to be exchanged in diffractive processes.

\section*{GENERIC DEFINITION OF DIFFRACTION}
The generic signature of diffraction, which applies to all soft 
and hard diffractive processes, is the presence of one (or more) 
rapidity gap(s) in an event, whose probability of formation is not 
exponentially damped. Generally, the exchange of a gluon, a quark or 
a color-singlet particle state, such as a $\rho$-meson, between two protons 
at high energies leads to events in which, in addition to whatever 
hard scattering may have occured, the entire rapidity space is 
filled with soft (low momentum) particles 
{\em (underlying event)}. 
The soft particle distribution is approximately flat in 
rapidity.
The flat $dN/dY$ shape is the result of the $x$-scaling~\cite{Feynman} of  
the parton distribution functions of the incoming protons.
The total particle multiplicity is given by 
$N(s)=\int \rho\, dY=\rho\ln s$, 
where $\rho$ is the average 
particle density in rapidity space and $s$ is in GeV$^2$. 
Elastic scattering occurs through the 
Poisson fluctuation of the multiplicity to $N=0$ and therefore is 
$\sim e^{-\rho \ln s}\sim 1/s^{\rho}$. Thus, on general 
field-theoretical grounds, the elastic scattering cross section 
is expected to fall with increasing energy for any exchange 
that has quantum numbers other than those of the vacuum, 
since the acceleration of the field associated with the quantum numbers 
of the exchange produces radiation that results in a positive particle density 
$\rho$. The experimental finding that the elastic cross section 
at high energies not only does not fall but actually increases with energy,
led to postulating a new kind of an exchange, the {\em pomeron}, 
defined as a state with the quantum numbers of the vacuum. 
Since no radiation is emitted by the acceleration of vacuum quantum numbers,
pomeron exchange leads to rapidity gaps whose probability is not damped 
exponentially. Therefore, such gaps 
are an unmistakable signature for diffractive production and 
can be considered as the {\em generic} definition of diffraction.  
In QCD, the pomeron can be thought of as a 
color-singlet state of quarks and gluons, whose structure can be 
probed in hard diffraction processes.
 
	Rapidity gaps can be formed in all inelastic  
non-diffractive (ND) events by multiplicity fluctuations. 
The probability for a gap of 
width $\Delta Y=Y-Y_{min}$ within a ND event sample is given by 
\begin{equation}
P^{ND}_{gap}(\Delta Y)=\rho\,e^{-\rho\,\Delta Y}
\label{NDG}
\end{equation}
This probability is normalized so that its integral is unity. 
A similar expression can be written for 
the {\em generic} gap probability distribution in single-diffraction
dissociation (SD),
\begin{equation}
P^{SD}_{gap}(\Delta Y)=K\cdot e^{n\,\Delta Y}
\label{SDG}
\end{equation}
Here, the parameter $n$ is positive and hence 
the gap probability grows with increasing $\Delta Y$, in accordance 
with our generic definition of diffraction. The actual gap probability 
distribution in a class of diffractive events will be the generic 
probability modulated by the cross section of the associated pomeron,
which generally depends on the width of the gap (see below).
The normalization factor, $K$, is obtained by setting 
the integral of the generic gap probability to unity.
Such a normalization yields 
$K=n/(e^{n\Delta Y_{max}}-1)$, which for $pp\rightarrow pX$, 
where $\Delta Y_{max}=\ln s$,  
becomes $K={n}/({s^n}-1)\approx n/s^n$. 
Note that the normalization depends only on $s$
and does not affect the shape of the gap probability.

A rapidity gap in SD is associated with a pomeron, which carries 
a fraction $\xi$ of the momentum of the proton.
The gap width, $\Delta Y$, is related to $\xi$ by 
\begin{equation}
\Delta Y=\ln\frac{1}{\xi}
\label{RG}
\end{equation}
In terms of $\xi$, Eq.~\ref{SDG} becomes 
\begin{equation}
P^{SD}(\xi)=K\cdot \frac{1}{\xi^{1+n}}
\label{SDGXI}
\end{equation}
Above, we have assumed 
that $n$ is independent of the 4-momentum transfer, $t$,
carried by the pomeron that forms the gap. 
We now assume, more generally, a linear dependence of $n$ on $t$,
which we parametrize as $n(t)\equiv 2(\epsilon+\alpha'\,t)$, and express $n(t)$ 
in terms of a new variable, $\alpha(t)$, defined as 
$\alpha(t)\equiv 1+\epsilon+\alpha'\,t$. Eq.~\ref{SDGXI} takes the form
\begin{equation}
P^{SD}(\xi,t)=K\cdot \frac{1}{\xi^{2\alpha(t)-1}}\cdot F^2(t)
\label{P-flux}
\end{equation}
where we have included the term $F^2(t)$, which represents the form 
factor of the proton, namely the probability that it remains 
intact after receiving a 
4-momentum transfer kick $t$. This formula has exactly the form of the pomeron 
flux in Regge theory~\cite{GM}, where $\alpha(t)$ is the pomeron trajectory, 
except for one important difference: since it represents the mapping into 
$(\xi,t)$-space of  
the {\em generic} gap probability distribution, 
it must be normalized so that its integral 
over all available phase space in $\xi$ and $t$ be unity. 
Such a (re)normalization of  the pomeron flux was proposed by this 
author~\cite{R} and its predictions have been shown to agree 
with data on soft~\cite{GM,KGF} and hard~\cite{KGF,KGDIS} diffraction.

\section*{RESULTS}
In this section we examine results on soft and hard diffraction 
and compare their main 
features with those expected from the generic definition of 
diffraction presented above. 
\subsection*{Soft Diffraction}
An analysis~\cite{GM} of $pp$ and $\bar pp$ single diffraction dissociation 
data has 
shown that, after subtracting the meson exchange contribution, the cross 
section can be expressed as a product of the generic gap probability, 
represented by Eq.~\ref{P-flux}, times a factor representing the 
$\pom-p$ total cross section, taken (from Regge theory) to be  
$\sigma^{\pom p}=\sigma_0^{\pom p}\cdot (s\xi)^{\epsilon}$. 
Note that the gap probability in the SD event sample is no longer 
given by Eq.~\ref{SDG}, since it is modulated by the $\xi^{\epsilon}=
e^{-\epsilon\Delta Y}$ dependence of the $\pom-p$ total cross section.
\subsection*{Hard Diffraction at HERA}
At HERA, both the H1 and ZEUS Collaborations used deep inelastic 
scattering (DIS) to measure the ``diffractive 
structure function" of the proton, 
$F_2^{D(3)}(Q^2,\beta,\xi)$ (integrated over $t$), where $\beta$ is 
the fraction of the momentum of the pomeron taken by the struck quark.
Both experiments found the form
\begin{equation}
F_2^{D(3)}(Q^2,\beta,\xi)=\frac{1}{\xi^{1+n}}\cdot A(Q^2,\beta)
\label{F2D3}
\end{equation}
in which the variable $\xi$ factorizes out into an expression
that has the $\xi$-dependence of the standard Regge theory pomeron flux factor.
Therefore, it appeared reasonable to  
consider the term $A(Q^2,\beta)$ as being proportional to 
the pomeron structure function $F_2^{\pom}(Q^2,\beta)$. 
The $A(Q^2,\beta)$ term was found to be rather flat in $\beta$,
suggesting that the pomeron has a {\em hard} quark structure. 
For a fixed $\beta$, $A(Q^2,\beta)$ increases with $Q^2$. 
By interpreting the $Q^2$ dependence to be due to scaling violations, 
the H1 Collaboration extracted the gluon fraction of the pomeron 
using the DGLAP evolution equations in a QCD analysis of 
$F_2^{D(3)}(Q^2,\beta,\xi)$.
The ZEUS Collaboration determined the gluon fraction by combining 
information from diffractive DIS, which is sensitive mainly to the 
quark component of the pomeron, 
and diffractive dijet photoproduction, which is 
sensitive both to the quark and gluon contents.
Both experiments agree that the pomeron structure is hard and consists of 
gluons and quarks in a ratio of approximately 3 to 1.
The extracted 
gluon fraction does not depend on the normalization of the 
$F_2^{D(3)}(Q^2,\beta,\xi)$ (for H1) or on the normalization of the 
pomeron flux (for ZEUS). 

Addressing now the question of normalization, if the $1/\xi^{1+n}$ term 
in Eq.~\ref{F2D3} 
represented the generic gap probability, rather than the 
standard pomeron flux, it should be normalized so 
that its integral from $\xi_{min}=Q^2/\beta s$ to $\xi_{max}=1$  be unity. 
Such a normalization yields $K=n(Q^2/\beta s)^n$ and therefore 
$F_2^{D(3)}$ can be 
written as 
\begin{equation}
F_2^{D(3)}(Q^2,\beta,\xi)
\sim \left[\left(\frac{Q^2}{\beta s}\right)^n\cdot \frac{1}{\xi^{1+n}}\right]
\cdot F_2^{\pom}(Q^2,\beta)
\label{F2D3P}
\end{equation}
where the term in brackets represents the generic $\xi$-probability.
Since the measured $Q^2$ dependence of 
$F_2^{D(3)}$ is represented well~\cite{R} by the factor $(Q^2)^n$, 
which belongs to the (pomeron flux) 
term in the brackets, the pomeron structure, 
$F_2^{\pom}(Q^2,\beta)$, must be largely independent of $Q^2$.
The asymptotic gluon momentum fractions for a quark-gluon construct 
for $2\div 3\div 4$ quark flavors is $0.73\div 0.64\div 0.57$~\cite{R}.
The gluon fraction found by ZEUS, $0.3<f_g<0.8$, agrees with 
the fraction expected from the asymptotic rules. The fraction found 
by H1 is not relevant if the $Q^2$ dependence does not belong to 
the pomeron structure. 
\subsection*{Hard Diffraction at the Tevatron}
Both the CDF and D\O\, Collaborations have reported 
that the jet $E_T$ distributions from non-diffractive (ND), single 
diffractive (SD) and double pomeron exchange (DPE) dijet events 
have approximately the same shape~\cite{DJJ,DPE,DPED0}. 
Since in going from ND to SD or from SD to DPE 
a nucleon of momentum $p$ is replaced by a pomeron of momentum $p\xi$,  
the similarity of the $E_T$ spectra 
suggests that the pomeron structure must be harder than the 
structure of the nucleon by a factor of $\sim 1/\xi$.
Assuming a hard pomeron structure, the CDF Collaboration 
determined the gluon fraction of the pomeron to be $f_g=0.7\pm 0.2$ 
by comparing the measured rate of diffractive $W$ production, 
which is sensitive to the quark content of the pomeron, with the rate for 
diffractive dijet production, 
which depends on both the quark and gluon contents.
The measured gluon fraction, which is independent of the pomeron flux 
normalization assumed in the Monte Carlo simulations,
agrees with the result obtained by ZEUS.

For a hard pomeron structure with $f_g=0.7$ and $f_q=0.3$, 
the measured $W$ and dijet rates are smaller than the rates 
calculated using the standard pomeron flux by a factor $D=0.18\pm 0.04$.
This flux ``discrepancy" factor is consistent with the pomeron flux 
renormalization expectation~\cite{R,KGDIS,WJJ} and therefore consistent 
with the generic $\xi$-distribution probability. 

The CDF Collaboration also measured the rate for DPE dijets and 
compared it with the rates for SD and ND dijets and with calculations 
using the standard pomeron flux. To obtain the measured DPE/SD ratio, 
the standard flux in DPE must be multiplied by the factor $D$ {\em 
for both the proton and antiproton}. This result 
supports the hypothesis that the suppression factor, relative to the 
standard flux calculations, is associated with the 
normalization of the $\xi$ probability distribution, rather than 
with  ``screening corrections" as proposed by other authors~\cite{GLM,K2}.
\subsection*{From HERA to the Tevatron}
The rate for diffractive $W$ production at the Tevatron 
can be calculated directly from 
$F_2^{D(3)}(Q^2,\beta,\xi)$~\cite{KGDIS,Whitmore}.
Using conventional factorization, the expected SD to ND ratio for $W$ 
production is $\sim 7\%$.
By scaling the $F_2^{D(3)}(Q^2,\beta,\xi)$ by the normalization 
factors of the $1/\xi^{1+n}$ term at HERA, where $\xi_{min}=Q^2/\beta s$, 
to that at the Tevatron, where $\xi_{min}=M_0^2/\beta s$ with 
$M_0^2\approx 1.5$ GeV$^2$, 
the prediction 
becomes 1.24\%~\cite{KGDIS}, which agrees with the data. This result 
supports the hypothesis of an underlying generic $\xi$ probability 
distribution given by Eq.~\ref{SDGXI}.

\section*{CONCLUSIONS}
We have presented a generic definition of diffraction, which is  
based on the formation of rapidity gaps that are not exponentially damped.
The formation of a diffractive rapidity gap is presumed to be associated 
with the exchange of a pomeron, defined as a color-singlet state 
with the quantum numbers of the vacuum. The quark-gluon 
structure of the pomeron can be probed in diffractive processes 
that incorporate a hard scattering (hard diffraction).
Results on hard diffraction from
HERA and from the Tevatron indicate that the pomeron structure is hard 
and consists of gluons and quarks in a ratio of approximately $3\div 1$.
A comparison of hard diffraction rates at HERA 
with rates at the Tevatron confirms the hypothesis embedded in the
generic definition of diffraction that the gap probability 
must be normalized to unity, i.e. scaled to its integral over all available 
phase space for gap formation. The scaling of the gap probability 
violates conventional factorization, but 
respects unitarity and leads to an unambiguous normalization.
Moving beyond phenomenology and providing a QCD-based picture of 
the pomeron that can explain the experimental results is clearly the next  
true theoretical challenge.

\end{document}